\documentclass[conference ]{IEEEtran}
\IEEEoverridecommandlockouts
\usepackage{cite}
\usepackage{amsmath,amssymb,amsfonts}
\usepackage{algorithmic}
\usepackage{graphicx}
\usepackage{textcomp}
\usepackage{xcolor}
\usepackage{float}
\usepackage{subfig}
\usepackage{mathtools}
\def\BibTeX{{\rm B\kern-.05em{\sc i\kern-.025em b}\kern-.08em
    T\kern-.1667em\lower.7ex\hbox{E}\kern-.125emX}}
\begin{document}

\title{Advancing Voice Cloning for Nepali: Leveraging Transfer Learning in a Low-Resource Language
 \\
}

\author{
 \begin{tabular}{ c c c }
Manjil Karki & Pratik Shakya & Sandesh Acharya\\
\textit{\normalsize Department of Computer Engineering} & \textit{ \normalsize Department of Computer Engineering} & \textit{ \normalsize Department of Computer Engineering}\\
\textit{\normalsize Khwopa College Of Engineering} &\textit{\normalsize Khwopa College Of Engineering}&\textit{\normalsize Khwopa College Of Engineering}\\
\normalsize
Bhaktapur, Nepal & \normalsize Lalitpur, Nepal & \normalsize Lalitpur, Nepal \\
\normalsize KCE075BCT019@khwopa.edu.np & \normalsize KCE075BCT024@khwopa.edu.np & \normalsize KCE075BCT040@khwopa.edu.np \\
\end{tabular}\\
\and
\hspace{3cm}
\begin{tabular}{ c c }
Ravi Pandit & Dinesh Gothe\\
\textit{\normalsize Department of Computer Engineering} & \textit{ \normalsize Department of Computer Engineering}\\
\textit{\normalsize Khwopa College Of Engineering} & \textit{\normalsize Khwopa College Of Engineering}\\
\normalsize Bhaktapur, Nepal&\normalsize Bhaktapur, Nepal  \\
\normalsize KCE075BCT028@khwopa.edu.np & \normalsize gothe.dinesh@khwopa.edu.np\\
\end{tabular}
}
\maketitle

\begin{abstract}
Voice cloning is a prominent feature in personalized speech interfaces. A neural vocal cloning system can mimic someone's voice using just a few audio samples. Both speaker encoding and speaker adaptation are topics of research in the field of voice cloning. Speaker adaptation relies on fine-tuning a multi-speaker generative model, which involves training a separate model to infer a new speaker embedding used for speaker encoding. Both methods can achieve excellent performance, even with a small number of cloning audios, in terms of the speech's naturalness and similarity to the original speaker.
Speaker encoding approaches are more appropriate for low-resource deployment since they require significantly less memory and have a faster cloning time than speaker adaption, which can offer slightly greater naturalness and similarity. The main goal is to create a vocal cloning system that produces audio output with a Nepali accent or that sounds like Nepali. For the further advancement of TTS, the idea of transfer learning was effectively used to address several issues that were encountered in the development of this system, including the poor audio quality and the lack of available data.
\end{abstract}

\section{INTRODUCTION}
Voice cloning involves creating artificial replicas of human voices using advanced AI software, sometimes indistinguishable from real voices. It's often associated with terms like deepfake voice, speech synthesis, and synthetic voice. Unlike text-to-speech (TTS) systems, which transform written text into speech using predefined data, voice cloning is a customized process. It extracts and applies specific voice characteristics to different speech patterns. Historically, TTS had two approaches: Concatenative TTS, which used recorded audio but lacked emotion, and Parametric TTS, which used statistical models but produced less human-like results. Today, AI and Deep Learning are improving synthetic speech quality, making TTS applications widespread, from phone-based systems to virtual assistants like Siri and Alexa.

\section{Methodology}
\subsection{Proposed Method:}
\begin{figure}[H]
    \centering
    \includegraphics[width=0.55\textwidth]{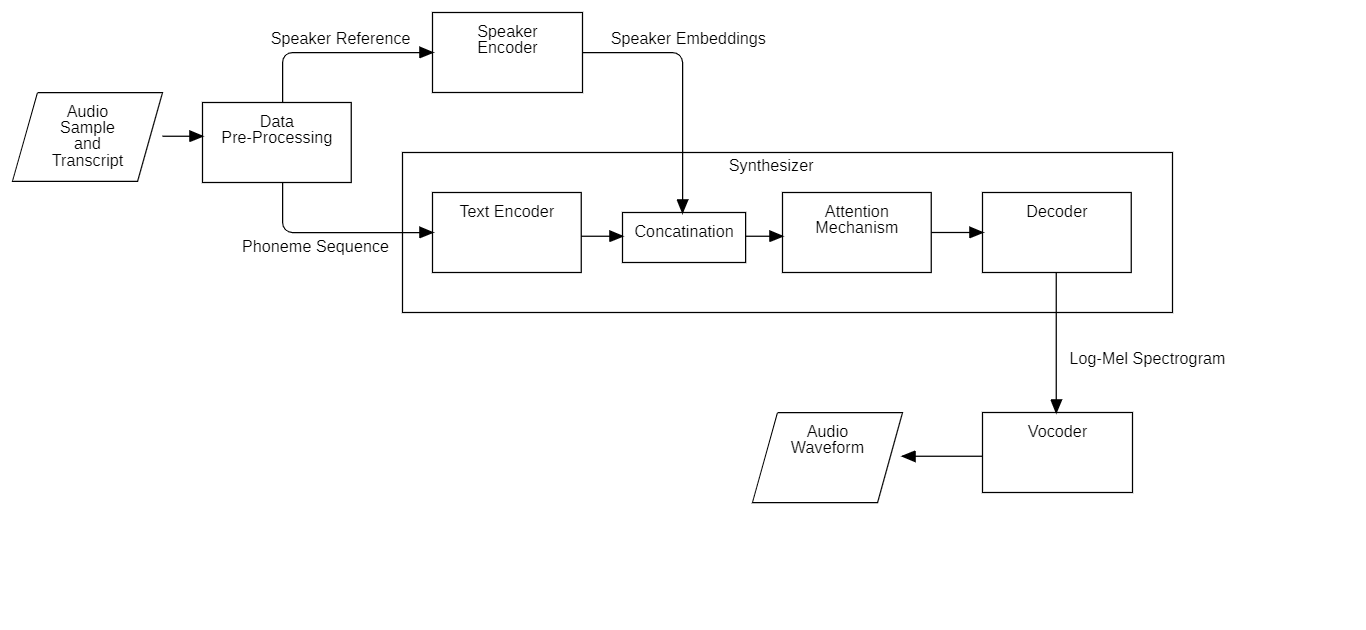}
    \caption{Block Diagram Nepali Voice Cloning}
    \label{fig:blockdiagram}
\end{figure}
The proposed method of voice cloning consists of three major models with initial preprocessing of data files \cite{b1}.
\subsection{Nepali Speech Corpus Creation} 
A \textless voice, text\textgreater pair multispeaker dataset is prepared consisting of 546 individuals cooperating with both male and female speakers. The dataset consists of female speakers majority. The dataset was taken from open source platform OpenSLR\cite{b2}\cite{b3}. The dataset consists of 150,000 plus audio files totalling up to 168 hours of audio with their respective transcript.
\begin{figure}[H]
    \centering
    \includegraphics[width=0.30\textwidth]{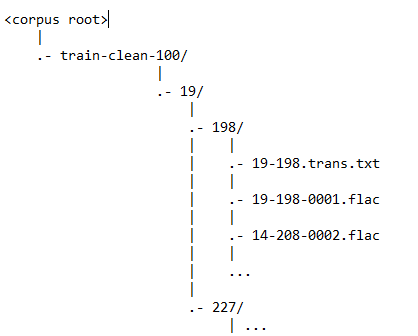}
    \caption{Structure of speech Corpus}
    \label{fig:dataset_corpus}
\end{figure}

\subsection{Data-preprocessing}
\begin{figure}[H]
    \centering
    \includegraphics[width=0.35\textwidth]{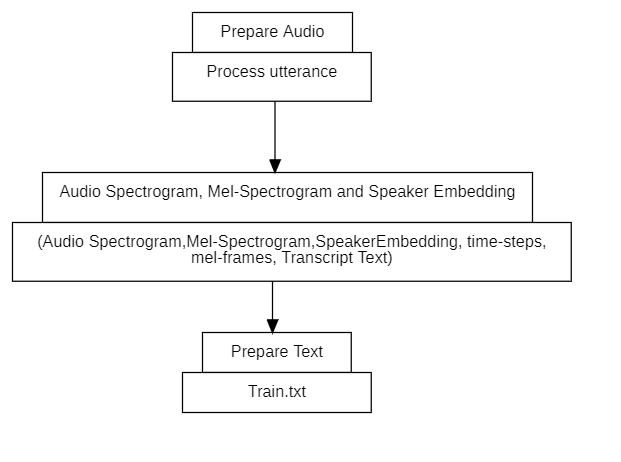}
    \caption{Preprocessing step for synthesizer model}
    \label{fig:preprocessing}
\end{figure}

\begin{figure}[H]
    \centering
    \includegraphics[width=0.50\textwidth]{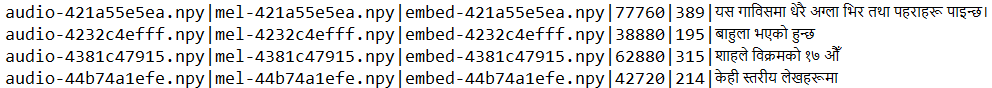}
    \caption{Example content of train.txt }
    \label{fig:trainText}
\end{figure}

Data pre-processing is a crucial step as the 3 models require a separate pre-processing. Initially for encoder-preprocessing the audio files from the dataset are fetched and processed so that an encoded form of mel-spectrogram is gained. Then for the synthesizer preprocessing, audio files along with transcript and utterance are fetched that are processed representing a preprocessed data collection including audio-spectrogram, mel-spectrogram, speaker-embedding and also a text file including the details. Finally, for the vocoder-preprocessing the mel-spectrograms were processed to ground-truth aligned (GTA) spectrum dataset.\\
Fig \ref{fig:preprocessing} shows the overview of the pre-processing step. Text
data will be well normalized and then audio files will be converted into both mel-scale spectrograms and Speaker Embeddings. Fig. \ref{fig:trainText} shows the contents from training data as an example.

\subsection{Encoder:}
The encoder is responsible for extracting meaningful representations (embeddings) from the input voice samples. It captures the characteristics and features of the speaker's voice and encodes them into a fixed-length vector. The architecture of the encoder can vary, but one popular choice is the Mel-spectrogram-based encoder\cite{b4}. The Mel-spectrogram-based encoder typically consists of several convolutional layers followed by batch normalization and non-linear activation functions, such as ReLU. It takes the raw audio waveform as input and converts it into a Mel-spectrogram, which is a visual representation of the audio's frequency content over time. The Mel-spectrogram is then passed through the convolutional layers to extract features, which are finally transformed into an embedding vector using pooling or recurrent neural networks (RNNs).

\subsection{Sythensizer:}
Tacotron2\cite{b5} is a popular architecture for text-to-speech synthesis. It consists of a text encoder that converts input text into a fixed-length embedding and a decoder that generates mel-spectrograms representing the desired speech. The text encoder uses convolutional layers and bidirectional recurrent neural networks (RNNs) to capture text features. The decoder utilizes autoregressive models with stacked LSTM\cite{b19} or GRU layers, along with attention mechanisms, to generate mel-spectrograms. During training, Tacotron2 minimizes a loss function to make the generated spectrograms match the target speech representation. It has been successful in producing high-quality synthetic speech.

\subsubsection{Architecture}

The Tacotron text-to-speech (TTS) system employs a deep neural network architecture comprising multiple components that work collaboratively to generate speech from text.

The key components of Tacotron2's architecture include:
\begin{itemize}
    
    \item Text Encoder: The initial component is the text encoder, responsible for converting a sequence of characters or phonemes into hidden representations. Typically, convolutional layers are employed to extract text features, followed by a bidirectional recurrent neural network (RNN) that encodes the input sequence into a fixed-length vector. These hidden representations capture the semantic meaning of the text and serve as input for the decoder component.

    \item Attention Mechanism: The attention mechanism is a vital component in Tacotron\cite{b6}. It learns to align the output spectrograms with the input text sequence. By computing attention weights, it determines the importance of each input sequence element at each decoder time step. These weights are used to calculate a weighted sum of the encoder's hidden representations, which becomes input for the decoder.

    \item Decoder: The decoder makes a sequence of speech spectrograms, which represent the acoustic features of the speech waveform. It typically consists of a recurrent neural network with an attention mechanism. The decoder generates one spectrogram frame at a time, using the previous frame along with current decoder's hidden state. The attention weights and the previous decoder output are utilized as inputs for generating the subsequent spectrogram frame. The decoder's output undergoes post-processing through a network to obtain the final spectrogram.
    
    \item Post-Processing Network: The decoder's output is a sequence of speech spectrograms that capture the acoustic features of the speech waveform. These spectrograms are transformed into a speech waveform using techniques such as Griffin-Lim phase reconstruction. This involves estimating the phase information of the waveform from the magnitude spectrogram produced by the decoder. The resulting waveform then passes through a post-processing network, which applies additional signal-processing steps to enhance the quality of the synthesized speech.
    
\end{itemize}

\subsubsection{Alignment Plots}

Alignment plots are visualizations used in speech processing and natural language processing. They show the alignment between two sequences, such as audio and its transcription. They help evaluate the performance of automatic speech recognition or machine translation systems\cite{b7}. The plots consist of parallel sequences displayed on the x-axis and y-axis, representing the original audio and predicted transcription/translation. Each cell represents the alignment between specific segments. Alignment plots can be created using heat maps, scatterplots, or line plots. They allow for visual comparison, error identification, and system improvement. Alignment plots are valuable for understanding system performance in speech and language processing.

\subsubsection{Mel-Spectrogram}

A Mel spectrogram\cite{b8} is a visual representation of the power spectrum of an audio signal, where the frequencies are weighted according to the Mel scale. It is widely used in speech and audio processing for tasks like speech recognition and music analysis. The process involves dividing the audio signal into short segments, computing the power spectrum using a Fourier transform, and grouping frequencies into triangular bands using a filter bank. The power within each band is summed up to create Mel spectrogram coefficients. Mel spectrograms provide a powerful way to visualize and analyze the spectral content of audio signals over time. They are also used as input features for machine learning models in various audio processing tasks. Mel spectrograms emphasize important frequency bands while reducing less significant ones, helping to preserve relevant information while reducing dimensionality.

\subsection{Vocoder:}
WaveNet is a popular vocoder architecture used in voice cloning \cite{b9}. It employs dilated convolutional neural networks (CNNs) to model the conditional probability distribution of the audio waveform. WaveNet takes mel-spectrograms as input and generates the corresponding high-quality audio waveform sample by sample. It captures long-range dependencies in the audio using stacked dilated convolutional layers. WaveNet has been widely adopted for its ability to generate realistic and natural-sounding speech.

\subsection{Transfer Learning:}
Transfer learning aims to transfer knowledge from the source domain to the target domain to solve the problem caused by
insufficient training data and improve the generalization ability of
the model\cite{b17}. During the development of deep learning models for a specific architecture, initial attempts involved training the models from scratch. However, these models yielded poor results, primarily due to limited and low-quality data. Moreover, the training process itself was complex and time-consuming, especially on advanced systems. To address these issues, a solution was proposed and implemented: transfer learning for all three models. However, finding a language model that closely resembled the Nepali language and the voice of Nepali speakers proved impossible. As an alternative, a multi-language model was chosen for transfer learning. Surprisingly, the results obtained were significantly better than those of the previously trained models, although they still required some fine-tuning. Eventually, the necessary adjustments were made, resulting in somewhat improved outcomes.

\section{Result}

\subsection{Training Analysis}
The results that were observed during training are mostly the charts and plots which show the development of the model. Mostly the loss curve and accuracy curves are key parameters to watch for during training but this model is an audio-based model which is why subjective observations are given more emphasis as accuracy and losses could not cover all areas of interest.

\subsection{Encoder Training}
For encoder training some of the key parameters that were observed are listed as follows:

\subsubsection{Encoder Training Loss}
The encoder training loss is represented by the general end-to-end loss that states the accuracy of cluster formation in UMAP. 
The GE2E loss simulates this process to optimize the model. At training time, the model computes the embeddings $e_{i j} (1 \leq I \leq N, 1 \leq j \leq M)$ of M utterances of fixed duration from N speakers. A speaker embedding $c_{i}$ is derived for each speaker:
\begin{equation}
    C_{i}=\frac{1}{M}\sum_{j=1}^{M}e_{i j}.
\end{equation}\\

\subsubsection{Encoder Equal Error Rate}
Equal Error Rate (EER) is a performance metric used in biometric verification systems, which measures the similarity between two biometric samples. The EER is the point at which the False Acceptance Rate (FAR) equals the False Rejection Rate (FRR). The FAR is the proportion of impostor samples that are incorrectly accepted, while the FRR is the proportion of genuine samples that are incorrectly rejected. EER also represents the accuracy of the model in terms of similarity. The lower the value of EER more is the accuracy of the model. \\

Table \ref{tab:encoder_train} represents the result of encoder training.

\begin{table}[H]
        \centering
        \caption{Encoder Training}
        \begin{tabular}{|l|l|}
        \hline
        \multicolumn{2}{|c|}{ Encoder Training Result}  \\
        \hline
         Encoder Training Loss & $0.02 \pm 0.01$  \\
         \hline
         Equal Error Rate & $0.005 \pm 0.001$\\
         \hline
        \end{tabular}
        \label{tab:encoder_train}
    \end{table}
\subsubsection{UMAP Projection}
UMAP (Uniform Manifold Approximation and Projection)\cite{b12} is a dimensionality reduction technique used for visualization and clustering of high-dimensional data. It is particularly useful for nonlinear dimensionality reduction, meaning it can effectively handle data with complex patterns and nonlinear relationships between features. The output of UMAP is a two-dimensional projection of the data that can be plotted and visualized. UMAP is effective for a wide range of applications, including image analysis, genomics, and natural language processing.

\begin{figure}[H]
    \centering
    \subfloat[\centering Initial Cluster]{{\includegraphics[width=4.5cm]{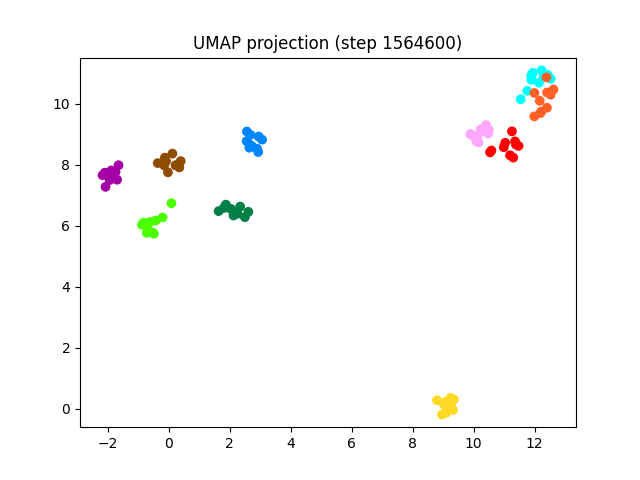} }}
    \subfloat[\centering Cluster after training]{{\includegraphics[width=4.5cm]{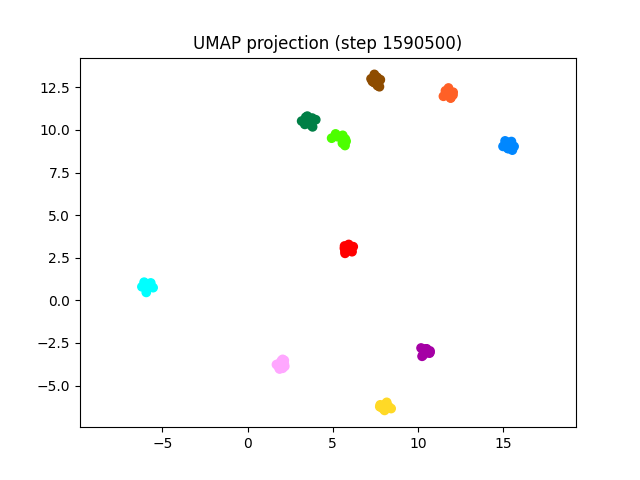} }}
    \label{fig:Umap}
    \caption{U-Map projection at different stages of training}
\end{figure}

\subsection{Synthesizer Training}
\subsubsection{M1 Loss}
M1 loss, also known as mean absolute error (MAE) loss, is a commonly used loss function in regression problems. It measures the average absolute difference between the predicted and actual values. The formula for MAE is:

\begin{equation}
    MAE = 1/n * \sum_{i=1}^n (yi - \hat{y}i)
\end{equation}

Where $yi$ is the actual value, $\hat{y}i$ is the predicted value, and n is the total number of samples.\\

\subsubsection{M2 Loss}
M2 loss, also known as mean squared error (MSE) loss, is another commonly used loss function in regression problems. It measures the average of the squared differences between the predicted and actual values. The formula for MSE is:
\begin{equation}
MSE = 1/n * \sum_{i=1}^n (yi - \hat{y}i)^2  
\end{equation}

Where $yi$ is the actual value, $\hat{y}i$ is the predicted value, and n is the total number of sample\\

\subsubsection{Final Loss}
The addition of M1 loss (mean absolute error) and M2 loss (mean squared error) is a method used in some regression problems to combine the strengths of both loss functions.M1 loss measures the average absolute difference between the predicted and actual values of a continuous variable, while M2 loss measures the average squared difference between the predicted and actual values. M1 loss is more robust to outliers, while M2 loss is more sensitive to the magnitude of errors. By combining these two loss functions, the resulting loss function can capture both the magnitude and direction of errors in the model's predictions. The addition of M1 and M2 loss can help the model balance the trade-off between accuracy and robustness and can improve its ability to generalize to new data.
\begin{equation}
Loss = MAE + MSE
\end{equation}
Table \ref{tab:synthesizer_train} represents the result of synthesizer training.

\begin{table}[H]
        \centering
        \caption{Synthesizer Training}
        \begin{tabular}{|l|l|}
        \hline
        \multicolumn{2}{|c|}{ Synthesizer Training Result}  \\
        \hline
         Mean Absolute Error & $0.293 \pm 0.001$  \\
         \hline
         Mean Squared Error & $0.0495 \pm 0.0001$\\
         \hline
         Total loss (MAE + MSE) & $0.3525 \pm 0.01$\\
         \hline
        \end{tabular}
        \label{tab:synthesizer_train}
    \end{table}
\subsubsection{Mel-spectogram}
Fig 6 shows the target and predicted mel-spectrogram during training.
\begin{figure}[H]
    \centering
    \includegraphics[width=0.45\textwidth]{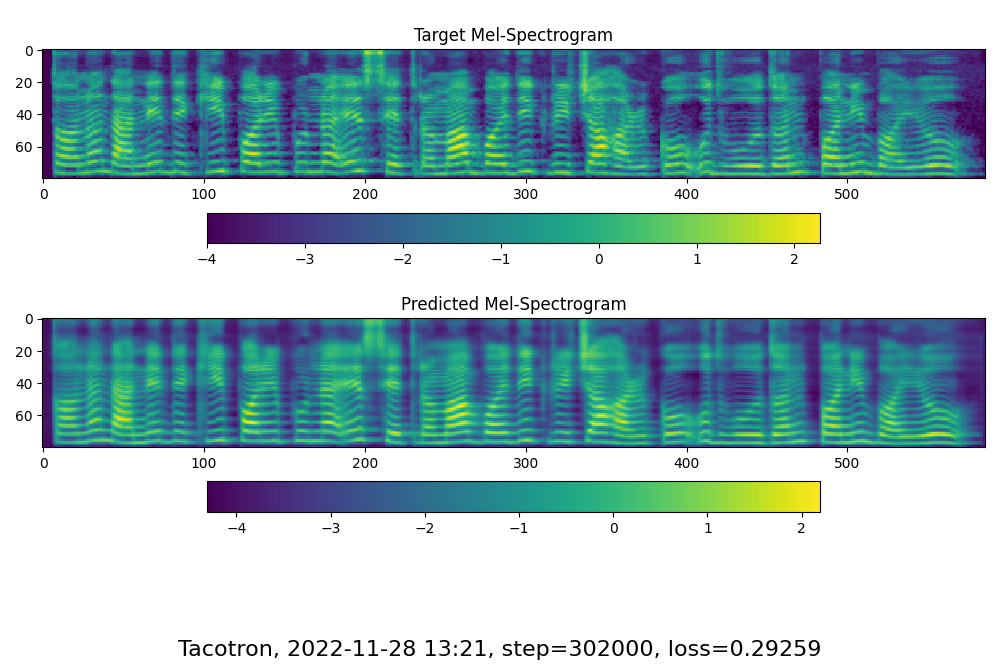}
    \label{fig:mel}
    \caption{Generated mel-spectrogram by the model during training}
\end{figure}  

\subsubsection{Alignment}
Fig 7 shows the learned alignment by attention mechanism.
\begin{figure}[H]
    \centering
    \includegraphics[width=0.3\textwidth]{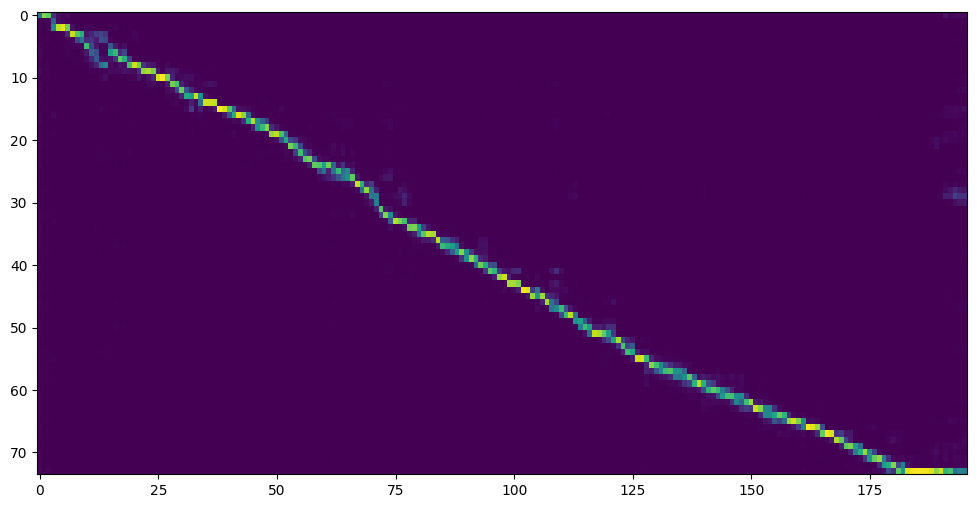}
    \label{fig:atten}
    \caption{Alignment plot of synthesizer model during training}
\end{figure}   
\subsection{Vocoder}

The vocoder training is done from the synthesizer audios and the GTA synthesized meals. Table \ref{tab:Vocoder_train} represents the result of this training.
\begin{table}[H]
        \centering
        \caption{Vocoder Training}
        \begin{tabular}{|l|l|}
        \hline
        \multicolumn{2}{|c|}{ Vocoder Training Result}  \\
        \hline
         Vocoder Training Loss & $4.095 \pm 0.005$  \\
         \hline
        \end{tabular}
        \label{tab:Vocoder_train}
    \end{table}
\subsection{Metrices}
Various tools, techniques and metrics were used while training our model for analysis and selection. Similar other metrics and methodologies were employed to verify and validate the results and final models.
\subsubsection{MOS score}
MOS (Mean Opinion Score) is a commonly used metric to evaluate the quality of audio or video signals, often in the context of telecommunications or multimedia applications \cite{b10}. It is a subjective assessment of the quality of the signal as perceived by human listeners or viewers. In a MOS test, a group of human subjects are asked to listen to or watch a set of audio or video samples, and rate the quality of each sample on a scale from 1 to 5 (or sometimes 1 to 10). The ratings are then averaged to obtain a MOS score for each sample. A higher MOS score indicates better quality, while a lower MOS score indicates poorer quality.\\
For MOS, subjective opinion mining of around 124 people was done where around 50 people had an idea about what the project was and the other 70 had no idea about the project. For collecting data a Google form was sent which had 10-10 audio samples of cloned and real voices. The form had a scale to rate naturalness and similarity from 1 to 5. The data was stored in Google Sheets which were processed with the help of pandas and python
Initially, the data needed to be cleaned for which NaN values were removed in the rows with more than 6 NaNs and rows containing less than 6 NaNs were filled with mean. Due to this, the data shrank down to only 120 users than average for each audio and the average for naturalness and similarity was calculated and displayed as follows.
\begin{figure}[H]
    \centering
    \includegraphics[width=0.45\textwidth]{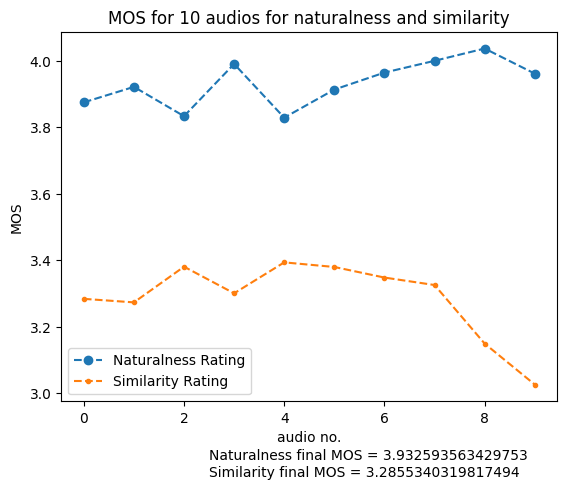}
    \caption{MOS(Mean Opinion Score)}
    \label{fig:MOS}
\end{figure}
Table \ref{tab:MOS_result} represents the MOS score for Naturalness and Similarity of the generated audio of 10 audio samples.
\begin{table}[H]
        \centering
        \caption{MOS representing audio Naturalness and Similarity}
        \begin{tabular}{|l|l|}
        \hline
        \multicolumn{2}{|c|}{ MOS Result}  \\
        \hline
        Audio Property & MOS  \\
         \hline\hline
         Naturalness & 3.93  \\
         \hline
         Similarity & 3.29  \\
         \hline
        \end{tabular}
        \label{tab:MOS_result}
    \end{table}
\subsubsection{U-MAP}
Similar to the map projection done in training the embeddings of the original voice and cloned voice were generated and after dimension reduction, the clusters were visualized. Fig 9 represents the UMAP plot where, in each cluster, the circles represent the original audio and the cross represents cloned audio. Close cluster formation of original and generated audio shows the greater accuracy of the models. 
\begin{figure}[H]
    \centering
    \includegraphics[width=0.25\textwidth]{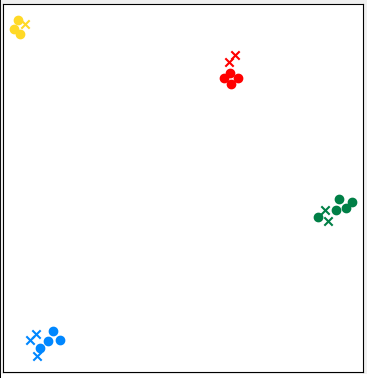}
    \caption{UMAP of multiple users and cloned audio}
    \label{fig:ump}
\end{figure}

\subsubsection{PESQ}
PESQ stands for Perceptual Evaluation of Speech Quality, which is a standard method for evaluating the perceived quality of speech in telecommunications systems\cite{b11}. PESQ is calculated by comparing a degraded speech signal to a reference, or "clean," speech signal. The degraded speech signal is typically generated by passing the reference signal through a telecommunications system or other communication channel that may introduce various types of distortion. The PESQ algorithm compares the degraded speech signal to the reference signal using a perceptual model that takes into account the characteristics of human hearing. The algorithm analyzes the signals in small segments and computes a quality score for each segment based on the similarity between the degraded and reference signals.\\

\noindent The PESQ algorithm produces a quality score for degraded speech signals that range from -0.5 to 4.5, with higher scores indicating better perceived quality. The score is typically expressed as a mean opinion score (MOS) ranging from 1 to 5, with 5 being the highest possible score and indicating "excellent" quality, and 1 being the lowest possible score and indicating "bad" quality.

\noindent In the case of our project we calculate PESQ for the cloned voice PESQ
for training dataset samples and test samples. We found it to be 2.8 in the case of validation data and 2.3 in the case of test data. The values we obtained were not so satisfying this is because of the quality of the training dataset.

\begin{figure}[H]
    \centering
    \includegraphics[width=0.35\textwidth]{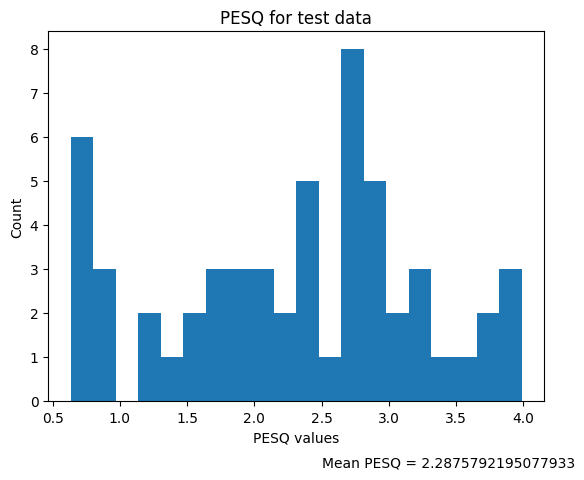}
    \caption{PESQ(Perceptual Evaluation of Speech Quality)}
    \label{fig:PESQ}
\end{figure}

\section{Conclusion}

 In conclusion, the system for Nepali Voice Cloning emphasizes cloning voices having Nepali accents, dialects etc taking in Devanagari script as input. For voice cloning the system has three core components namely: encoder, synthesizer and vocoder. The speaker Encoder consists of CNN followed by RNN that changes the User voice to speaker encoding, the encodings are used to distinguish a user from another it is a kind of fingerprint. The speaker embedding along with the text which is in Devanagari script is given in as input to the synthesizer. The synthesizer is further divided into three components (encoder, attention and decoder). It is a modification of the tacotron model which is capable of taking in speaker embeddings as input. The output from the synthesizer is mel-spectogram which is sent into the vocoder model to generate audio from mel-spectogram. Finally, all these components are integrated into a web application where the integrated system can run at its full potential.\\
 
 Mathematically, the system performs fair as it has an MOS score of 3.9 in terms of naturalness and an MOS score of 3.2 in terms of similarity on a scale of 1 to 5. Also, the mean PESQ score for validation data was 2.3 and for the test, the dataset was found to be 1.8. The scale of PSEQ ranges from -0.5 to 4.5.

\section*{Acknowledgment}

First and foremost, we would like to thank our primary supervisor, Er. Dinesh Gothe, for their guidance, support, and encouragement throughout the entire project. Also, we are deeply indebted to the staff members of the Computer department(Khwopa College of Engineering) for boosting our efforts and morale with their valuable advice and suggestions regarding the project and for supporting us in tackling various difficulties. Finally, we would like to thank Khwopa College of Engineering for providing the necessary hardware resources which helped us in the smooth progress of this research.\\

\end{document}